\documentclass[letterpaper,twoside,12pt]{article}
\usepackage{amssymb}
\usepackage{amsmath}
\usepackage{latexsym}
\usepackage{graphicx}
\usepackage{stmaryrd}
\usepackage[english]{babel}
\usepackage{setspace}
\usepackage[english]{babel}
\usepackage[utf8]{inputenc}
\usepackage[T1]{fontenc}
\usepackage{lmodern}
\usepackage{enumitem}
\begin{document}

%%%%%%%%%%%%%%%%%%%%%%%%%%%%%%%%%%%%%%%%%%%%%%%%%%%%%%%%%%%%%%
%%%%%%%%%%%%%%%%%%%%%%%%%%%%%%%%%%%%%%%%%%%%%%%%%%%%%%%%%%%%%%
\newcommand{\bra}[1]{\langle #1|}
\newcommand{\ket}[1]{|#1\rangle}
\newcommand{\braket}[2]{\langle #1|#2\rangle}
%%%%%%%%%%%%%%%%%%%%%%%%%%%%%%%%%%%%%%%%%%%%%%%%%%%%%%%%%%%%%%
%%%%%%%%%%%%%%%%%%%%%%%%%%%%%%%%%%%%%%%%%%%%%%%%%%%%%%%%%%%%%%

\begin{Large}
\begin{center}
\textbf{Losing stuff down a black hole}\\%[-5cm]
\end{center}
\end{Large}

\begin{center}
\begin{large}
Elias Okon\\
\end{large}
\textit{Instituto de Investigaciones Filos\'oficas, Universidad Nacional Aut\'onoma de M\'exico, Mexico City, Mexico.}\\[.5cm]
\begin{large}
Daniel Sudarsky\\
\end{large}
\textit{Instituto de Ciencias Nucleares, Universidad Nacional Aut\'onoma de M\'exico, Mexico City, Mexico.} \\
\end{center}

%\noindent
Over the years, the so-called \emph{black hole information loss paradox} has generated an amazingly diverse set of (often radical) proposals. However, forty years after the introduction of Hawking's radiation, there continues to be a debate regarding whether the effect does, in fact, lead to an actual problem. In this paper we try to clarify some aspect of the discussion by describing two possible perspectives regarding the landscape of the information loss issue. Moreover, we advance a fairly conservative point of view regarding the relation between evaporating black holes and the rest of physics, which leads us to advocate a generalized breakdown of unitarity. We conclude by exploring some implications of our proposal in relation with conservation laws. 

\onehalfspacing
%%%%%%%%%%%%%%%%%%%%%%%%%%%%%%%%%%%%%%%%%%%%%%%%%%%%%%%%%%%%%% 
\section{Introduction}
%%%%%%%%%%%%%%%%%%%%%%%%%%%%%%%%%%%%%%%%%%%%%%%%%%%%%%%%%%%%%%

Ever since the discovery by Stephen Hawking that black holes ought to emit thermal radiation, physicists have been trying to confront the rather uncomfortable implications of such a finding. There is, in fact, very little room for doubt about the calculation itself, relying, as is does, on a direct extrapolation of quantum field theory to the context of curved spacetimes, in complete accordance, among other important things, with the equivalence principle.\footnote{There is, though, one issue that is raised with some frequency: the trans-Planckian character of the modes involved. In our view, however, the mere formulation of such a concern is misguided simply because the is nothing intrinsically trans-Planckian about any mode in the absence of an ad hoc specification of a \emph{preferential} local inertial frame. In other words, any mode is trans-Planckian when characterized in a suitable local Lorentz frame. And, of course, the assignment of a privileged status to one such frame comes in directly contradiction local Lorentz invariance, a feature that underlies both general relativity and quantum field theory.} Moreover, there is an essentially universal consensus that, as the black hole radiates, its mass must decrease, leading to an increase in the radiation rate resulting in turn in a runaway process. However, at that point, the opinion of the community splits into an ever-diverging set of conflicting proposals, based on amazingly diverse perspectives and considerations: from arguments suggesting doors to other universes opening up deep in the black hole's interior to proposals regarding dramatic violations of the equivalence principle or large departures from the central tenets of quantum field theory, even in regions where the curvature is small. 
 
At any rate, even before all those proposals are to be examined, there is the question of whether there is indeed any problem at all to begin with. Is it not the case that a careful application of the relevant foundational principles leads to an unproblematic resolution of the issue? It is really the case that novel aspects of physics are to be inferred from the consideration of the black hole evaporation process? The aim of this paper is to explore these issues, trying to understand the main sources of disagreement regarding the plethora of answers to the above questions available in the literature. In addition, we analyze some of the implications of what we take to be fairly conservative views regarding these issues, taking seriously the question of how are they to be incorporated consistently with a unified view of all of physics.

In more detail, in section \ref{Two} we describe two important perspectives regarding the status and prospects of the information loss issue and in section \ref{now} we explore a proposal in \cite{Mau:17} regarding a novel foliation of an evaporating black hole spacetime. Next, section \ref{Mix} discusses issues regarding the interpretation of mixed states and their use in black hole scenarios and in section \ref{ADS-CFT} we make a few comments regarding the relation between the AdS/CFT correspondence and evaporating black holes. After that, in section \ref{qg} we evaluate the question of whether an evaporating black hole scenario may provide important clues for the construction of a quantum theory of gravity and in section \ref{BU} we discuss our own proposal which, as we will see in detail, calls for a generalized breakdown of unitarity.  Finally, in section \ref{con} we explore some important consequences of the framework we advocate regarding conservation laws and section \ref{C} closes with some brief concluding remarks.

%%%%%%%%%%%%%%%%%%%%%%%%%%%%%%%%%%%%%%%%%%%%%%%%%%%%%%%%%%%%%%
\section{Two perspectives on black hole evaporation}
\label{Two}
%%%%%%%%%%%%%%%%%%%%%%%%%%%%%%%%%%%%%%%%%%%%%%%%%%%%%%%%%%%%%%
Issues surrounding the so-called \emph{information loss paradox} continue to lead to heated discussions, which often take an acrimonious tone---particularly when the researchers involved have substantially different backgrounds. Often, though, there seems to be no actual disagreement over the results of the theoretical analysis. A case in point is \cite{Mau:17}, where Tim Maudlin explores the issue by meticulously outlining a fairly conservative, largely non-controversial picture on the subject, which results from straightforward considerations in general relativity and quantum field theory in curved spacetimes (together with very reasonable assumptions regarding the delicate issue of backreaction). The resulting picture of such an analysis is presented in Figure 1, which is Maudlin's version of the Penrose diagram of an evaporating black hole first presented in Hawking's famous 1975 paper.
\begin{figure}[h]
\centering
\includegraphics[height=8cm]{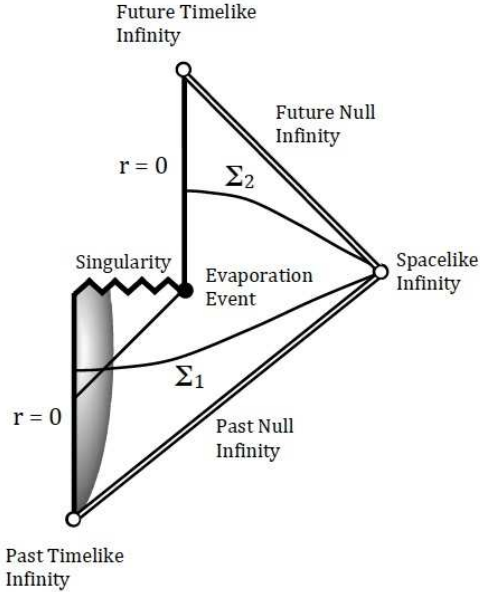} 
 \caption{Maudlin's version of an evaporating black hole Penrose diagram.}
\end{figure}

Maudlin's central observation is the indisputable fact that the hypersurface $\Sigma_2$ in Figure 1 is not a Cauchy hypersurface, from which it follows that one should not expect all the information contained in $\Sigma_1$ to also be contained in $\Sigma_2$ (Maudlin also introduces a novel foliation of the spacetime of Figure 1, of which we will have more to say in section \ref{now}). Maudlin then argues---convincingly, we believe---that \emph{in this scenario} there is no such thing as an information loss paradox. As a result, he concludes that there is nothing new to learn from the situation and that talk of the paradox simply ought to cease. However, for different reasons, not all physicists agree with Maudlin's conclusion, a fact which often leads to the discussions we alluded to at the beginning (see, e.g., \cite{Sab}). Without trying to offer a general analysis for the disagreement, we would like to describe two possible perspectives on the issue that could explain why, at least some (us, for example), could fully agree with the technical details of Maudlin's analysis but still disagree with his opinion that there is nothing new to learn from the issue.

We believe that an important source of disagreement is a difference in perspective regarding what the \emph{ultimate} physical characterization of a scenario involving an evaporating black hole ought to be. On the one hand, to those whose main background is in general relativity, or even quantum field theory in curved spacetimes, the picture that emerges from Maudlin's analysis (via tools such as \emph{domains of dependence} or \emph{Cauchy hypersurfaces}) seems complete, unassailable and, in a sense, entirely standard. That is, if the classical description of spacetime in terms of a 4-d manifold with a pseudo-Riemannian metric is taken as fundamental and if Figure 1 is taken as an accurate and exhaustive characterization of the physical situation at hand, then talk of information loss is simply misplaced. 

On the other hand, to those with an eye on quantum gravity, the picture presented by Maudlin does not seem to be of much help in capturing the true nature of the whole process. Sure, the diagram in Figure 1 seems to accommodate the fact that the black hole evaporates, but does so without really taking into consideration in detail issues such as backreaction, or quantum gravitational effects. Therefore, one should take the standard analysis as valuable information regarding a deeper level of description, but not as completely settling the issue.

The point is that people in this latter group do not envision the classical notion of spacetime, in terms of a 4-d manifold with a pseudo-Riemannian metric, as \emph{fundamental}. For them, the general relativistic notion of spacetime is only an effective characterization, which is valid in certain circumstances as an approximated description of what, at the fundamental level, is something of a rather different nature; and, of course, systems such as evaporating black holes are precisely the type of systems where the approximation is expected to break. Then, for this second group, scenarios such as evaporating black holes are thought to represent ideal playing grounds to explore where the classical description of spacetime breaks down and, as such, are considered ideal settings to gather clues regarding possible modifications of the classical picture arising form such a breakdown. In section \ref{qg} we will explore whether this expectation, of which Maudlin is highly skeptical, is reasonable or not. %Before doing so, we explore a peculiar proposal present in Maudlin's work.
 
%%%%%%%%%%%%%%%%%%%%%%%%%%%%%%%%%%%%%%%%%%%%%%%%%%%%%%%%%%%%%%
\section{About Maudlin's foliation}
\label{now}
%%%%%%%%%%%%%%%%%%%%%%%%%%%%%%%%%%%%%%%%%%%%%%%%%%%%%%%%%%%%%% 
As we mentioned in the previous section, in \cite{Mau:17} Maudlin introduces a novel Cauchy foliation of the spacetime describing the evaporation of a black hole (see Figure 2).
\begin{figure}[h]
\centering
\includegraphics[height=8cm]{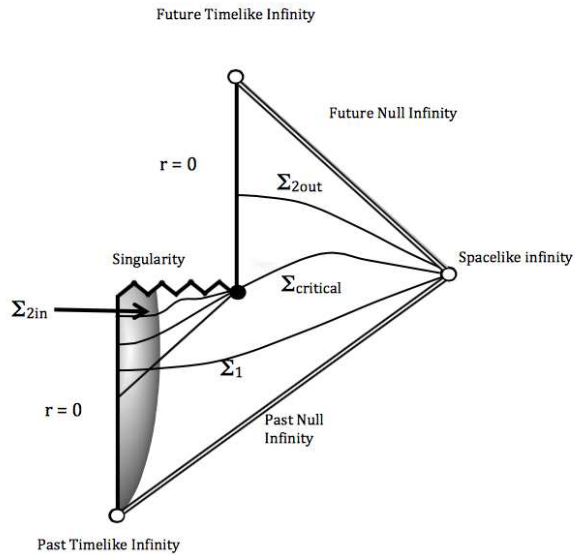} 
 \caption{Maudlin's diagram depicting the proposed foliation.}
\end{figure}
The idea is that before and up to the so-called Evaporation Event (the black dot in Figure 2), the Cauchy hypersurfaces are ``parallel'' to $\Sigma_1$, but that after that event there is no alternative but to continue the foliation by using \emph{disconnected} Cauchy surfaces such as $\Sigma_{2in} \cup \Sigma_{2out}$. Therefore, $\Sigma_{critical}$, which contains the Evaporation Event, divides the spacetime into an early region foliated by Cauchy surfaces that are connected and are topologically $R^3$ and a late region foliated by Cauchy hypersurfaces that are disconnected and topologically $R^3 \cup R^3$.\footnote{According to Maudlin, this change in topology evades Geroch's theorem because, due to the strange geometry at the Evaporation Event, the spacetime in question is not a manifold, and thus the global manifold structure presupposed in that theorem is not longer there (see \cite{Mau:17} for details).} 

Technically speaking, the construction seems to work just fine (except, perhaps, for issues related to the Evaporation Event and its strange geometric properties). Conceptually, though, one may be doubtful that it really is satisfactory, particularly given the conceptual role Cauchy hypersurfaces play regarding a generalized notion of simultaneity or what we normally refer to as ``now'' (i.e., the set of things that might be regarded as existing simultaneously with an event in which we find ourselves). In Newtonian scenarios, such a notion is of course straightforward. In relativistic contexts, where the concept has to be relativized and generalized, the standard practice is to use a (connected) Cauchy hypersurface for that purpose, and what Maudlin advances is that \emph{non-connected} Cauchy hypersurfaces can also serve the same purpose. However, one may ask if such a generalization really captures all the intuitive features one expects the concept of ``now'' to contain.

%For example, it is not clear if such a feature would survive a full resolution within some quantum gravity theory which cures the singularities. On the one hand a resolution of the singularity might be associated with the emergence of what is called a baby universe emerging on the other side of the quantum gravity region that replaces the singularity, a situation where one might indeed be force to contemplate a description which, at the semi-classical level, needs to be characterized as a topology change. On the other hand, one might have a situation where the curing of the singularity leads not to a baby universe but to a region of the same universe where asymptotic observers continue to live.

The fact is that it is not completely straightforward to identify what generates discomfort with Maudlin's proposal. One possible problem is that the non-connectivity feature implies that there are observers $O$ and $P$ on the same Cauchy hypersurface such that:
\begin{itemize}
\item There is no space-like geodesic connecting $O$ and $P$.%\footnote{As pointed out by Maudlin, in general within the general relativity, time-like connection between events is defined trough the use of time-like curves. In particular, one can often identify the geodesic between two time-like separated events as the time-like curve connecting the two events that maximizes the proper time (of course, sometimes there are more than one geodesic, but one can still identify the one which is longest and then refer to it as the ``maximizing geodesic''). On the other hand, space-like relationship between two events is often characterized simply by the absence of time-like curves between them. However, under normal circumstances, given two events that are space-like separated, one can identify the geodesic connecting them by a ``Mini-Max'' method as follows. Consider all open space-like hypersurfaces containing the two events; for each hypersurface consider the curves connecting the two events and restrict attention to hypersurfaces where the maxima is attained; choose that curve and the corresponding length and characterizing the hypersurface. From the collection of curves-hypersurfaces chose the one corresponding to the maximal length. The corresponding curve is a spacetime geodesic on the spacetime. Take the tangent vector at one event. Its projection on any hypesurface defines the direction on that hypersurface of the second event at the first event.)***REVISAR*** ME PARECE QUE LO PODRIAMOS SACAR : Si LO DEJAMOS HAY QUE AGRADECER A TONATIUH MIRAMONTES}
\item $O$ cannot even point in the direction of $P$.
\item $O$ cannot ever communicate with $P$.
\item $O$ cannot locate the ``obstacle'' that prevents his communication with $P$ (e.g. a horizon of some kind).
\end{itemize}

All this is in stark contrasts with ordinary situations where there are impediments to acquiring information about a certain event. When the spacetime contains, say, a particle horizon, making it physically impossible for an observer $O$ to ever acquire information about a particular event, at least the intersection of such a horizon with the hypersurface containing $O$ can be readily identified. As a result, given the instantaneous location of $O$, one can point towards the direction of the nearest intersection of the given hypersurface with such a horizon. Therefore, in the standard scenario with connected Cauchy hypersurfaces, even if there are obstructions to observers actually acquiring information about some particular event on that hypersurface, one is normally able to identify on the hypersurface the feature that constitutes the obstruction and point towards the direction in which it is located.

All this implies that, despite the fact that one might mathematically consider the extended notion of Cauchy hypersurafes described by Maudlin, and might therefore argue that, at late times, these non-connected Cauchy hypersurfaces are associated with quantum states that encode all the information that was present in the initial state, the situation seems rather anomalous. One would not be able, for instance, to characterize the direction in which certain observer $O$ would indicate that the missing information must be encoded. Of course, and as Maudlin clearly agues, information is a slippery concept (and more so when quantum theory is involved). In particular, the issue of \emph{locating} information is highly non-trivial, in which case the objection just presented seems to lose force. However, it turns out that a much more concrete notion that is also supposed to be conserved, namely, baryon number $B$, suffers a similar fate as information in the context of an eveporating black hole.\footnote{In fact, to be precise, one should consider $B-L$, baryon number minus lepton number, which is a quantity that is conserved both at the perturbative and non-perturbative levels in the Standard Model of particle physics. The conservation of $B$ itself is violated through highly suppressed ``sphaleron'' mediated processes. In the rest of the discussion we refer for simplicity to $B$ although, strictly speaking, we should be talking about $B-L$. \label{spha}} 

Consider, then, the formation of a black hole from the gravitational collapse of matter characterized by arbitrarily large $B$ (perhaps a Neutron star with $10^{50}$ baryons). When the black hole forms, given an hypersurface such as $\Sigma_1$ in Figure 2, one might compute $B$ in the exterior and associate with the intersection of such an hypersurface and the event horizon the $B$ that went into the black hole interior and argue that $B$ is indeed conserved. In fact, the association of the $B$ that went in with the boundary of the exterior region is nothing but a bookkeeping device to indicate that the part of the Cauchy hypersurface lying beyond the event horizon actually contains all the missing $B$. At any rate, any observer in the exterior of the black hole would be able to point towards the direction in which the missing $B$ is located.

Now, the evaporation of the black hole is essentially thermal (for most of the time and most of the energy emitted) and thus the radiation will contain as many baryons as anti-baryons, leading to a final state of the black hole exterior with essentially $B=0$ (or a very small number, certainly $\ll 10^{50}$). We might ask what happens with $B$ conservation once the black hole has evaporated completely. Well, extending Maudlin's approach to the $B$ question, one would say: nothing at all!, $B$ number is still $10^{50}$ and it is all concentrated on the part of the Cauchy surface that lies inside of the horizon. However, a collection of observers in the exterior part of the Cauchy hypersurface would have a hard time accepting this. They would note that, while they are supposed to agree that the $10^{50}$ baryons are ``present now,'' not only they cannot ever reach them, but they cannot even point in the general direction in which these baryons are supposed to be located, nor can they locate the feature that stops them from acquiring information about the missing baryons. That is, they end up with a notion of a \emph{now without a where}, and that is not what we normally think of as a \emph{now}. Under those circumstances, it seems that it would be more natural to say that the baryons simply do not exist at the time represented by the (connected component of the) hypersurface containing the observer in question. That is, at the \emph{effective present}, there are less baryons than there were before the gravitational collapse led to the formation of a black hole. In that case, one would be saying that $B$ is \emph{effectively} not conserved.

There is a further potentially serious complication with Maudlin's proposal we would like to mention. When we said his construction works fine at the technical level, we referred only to considerations within classical general relativity. However, in the context of quantum field theory in curved spacetimes (which is, of course, the context in which Hawking radiation is derived in the first place), a technical problem might arise. The issue, which is analogous to a component of the, so-called, firewall phenomenon \cite{firewalls}, is the following.\footnote{In \cite{Mau:17} it is argued that the firewall argument is not sound because it relies on the false assumption that $\Sigma_2$ is a Cauchy hypersurface. It very well might be the case that in \cite{firewalls} such an assumption is mistakenly held, but the firewall argument itself does not require it, it only requires the assumption that all the information is encoded in the outgoing radiation.} 

On the one hand, in Maudlin's scenario, the state of the field in $\Sigma_1$ is supposed to be unitarily mapped into a quantum state in $\Sigma_{2in} \cup \Sigma_{2out}$. This means that there must exist correlations between modes at points in $\Sigma_{2in}$ close to EE and points in $\Sigma_{2out}$ close to $r=0$. On the other hand, on any given hypersurface, special correlations between modes at nearby points are needed in order to ensure that the energy-momentum tensor of a quantum field does not involve uncontrollable divergences. In Maudlin's scenarios, this would require in particular correlations among modes at points in $\Sigma_{2out}$ close to $r=0$. The problem is that the, so-called, \emph{monogamy of entanglement} states that if maximal correlations are present between modes $A$ and $B$, then they cannot also be present between modes $A$ and $C$. Therefore, if the evolution is indeed unitary, the correlations among modes at points in $\Sigma_{2out}$ close to $r=0$ could not be there, so the energy-momentum tensor in $\Sigma_{2out} $ would have to exhibit uncontrollable divergences.\footnote{A full clarification of all these issues would requires a detailed technical analysis, something which goes beyond the scope of the present paper.}  Therefore, although the picture presented by Maudlin might be taken as viable within classical general relativity, it would seem to have to be discarded at the level of quantum field theory in curved spacetimes, unless a viable resolution of this problem is uncovered.
%%%%%%%%%%%%%%%%%%%%%%%%%%%%%%%%%%%%%%%%%%%%%%%%%%%%%%%%%%%%%% 
\section{Mixed states and black holes}
\label{Mix}
%%%%%%%%%%%%%%%%%%%%%%%%%%%%%%%%%%%%%%%%%%%%%%%%%%%%%%%%%%%%%%
The idea that, during the evaporation process, pure states evolve into mixed states, is commonplace in this type of discussions. However, what is not usually given much thought is the question of what is meant by the claim that the post-evaporation state is described by a mixed state. In this section we would like to discuss issues related to subtleties regarding the notion of a mixed state.

In general, mixed states have two different uses and one must be very careful as to which is appropriate in each situation. On the one hand, mixed states are used to describe either ensembles of identical systems, in which not all members of the ensemble are prepared in the same state, or situations in which one does not have full information regarding the actual pure state of the closed system one wants to study (and then relies on a probabilistic characterization of that state). These are the, so-called, \emph{proper mixtures} (see \cite{Espagnat} for the terminology). On the other hand, mixed states are also used to 
effectively describe a \emph{subsystem} of a closed quantum system which is itself, in a pure state. These are the, so-called, \emph{improper mixtures}. In the first case, the described systems possess at all times a well-defined quantum state, even though it might be unknown. In the second case, if the subsystem in question is entangled, it simply does not possess a well-defined pure quantum state.\footnote{Since different convex sums of pure states can correspond to the same mixed state, convex sums are better suited to encode epistemic uncertainty.} 

What about the possibility of assigning a \emph{mixed} state to a particular, isolated, non-entangled system, with the assumption that the lack of purity does not simply reflect the fact that we do not know the actual pure state of the system? That is, what about granting mixed states an ontological rather than an epistemic interpretation? Well, to begin with, it would not be clear what would that mean regarding the actual physical state of the system or the properties it possesses. The quantum mechanical rule to assign properties to a given system, namely, the Eigenvector-Eigenvalue rule,\footnote{The Eigenvector/Eigenvalue rule states that a physical system possesses the value $\lambda$ for a property represented by the operator $O$ if and only if the quantum state assigned to the system is an eigenstate of $O$ with eigenvalue $\lambda$.} would in this case lead to the claim that the system in question does not possess any properties whatsoever. 

Of course, the Eigenvector-Eigenvalue rule implies that most properties of systems described by \emph{pure states} do not possess well-defined values, but claiming instead that no properties whatsoever possess well-defined values is a different story. Such a claim could be reasonable if the system in question is an entangled subsystem, but it would be rather alarming if the system is assumed to be isolated and non-entangled to begin with (e.g., it would be rather strange to allow the theory to claim that the universe as a whole does not possess any properties).

Of course, one could try to come up with an alternative to the Eigenvector/Eigenvalue rule to try to solve this problem, but it is important to note that such a move, which is, by the way, never actually done, would take us far from standard quantum mechanics. Another possibility to solve this problem would be to adopt a purely instrumentalist position regarding quantum mechanics and forego the aspiration to describe the actual physical state of the system in question. That is, to take the quantum formalism as a mere tool to predict what we, human observers, would see. Well, this position may very well be tenable for standard, laboratory applications of quantum mechanics, but it is hardy adequate with non-standard applications of the theory, such as evaporating black holes, in mind. The problem, of course, is that the measurement problem, which already is problematic in the laboratory, is greatly exacerbated in situations in which observers and measuring apparatuses are not readily available.

Furthermore, it is a well-known property of mixed states, both proper and improper, that \emph{different} physical systems are assigned the \emph{same} mixed state.\footnote{Regarding proper mixtures, two ensembles of electrons, one with half of them spin-up and half of them spin-down along $z$ and the other with half spin-up and half spin-down, this time along $y$, are assigned the same density matrix. As for improper mixtures, a system of two electrons with state either $ \frac{1}{\sqrt2} \left( |+ \rangle^{(1)}_{z} |- \rangle^{(2)}_{z} + |-\rangle^{(1)}_{z} |+ \rangle^{(2)}_{z} \right)$  or $\frac{1}{\sqrt2} \left( |+ \rangle^{(1)}_{z} |+\rangle^{(2)}_{z} + |-\rangle^{(1)}_{z} |- \rangle^{(2)}_{z} \right)$ leads to the same reduced density matrix for, say, the second electron. It is interesting to note that all of the examples considered above lead to exactly the same density matrix; of course, this does not mean that the physical situations described are all equal (see \cite{P15} for a thorough discussion of this issue).}  This is a clear indication of the fact that the physical description provided by mixed states is \emph{incomplete}. It seems odd, then, to assume that mixed states can perform the same role as pure states as the description provided by the latter is assumed to be complete. Moreover, the possibility of assigning mixed states to isolated, non-entangled systems would signify an enormous explosion in the number of distinct physical states available to a system, which could have a negative impact on the thermodynamic properties of the system. Also, if such a position is adopted, it would be necessary, or at least desirable, to come up with a general principle indicating under exactly which physical conditions a state of a system would be either pure or mixed. Can one, for instance, assume without further analysis that a state of a system that leads to the formation of a black hole can be arranged to be initially pure? 

It is also interesting to note that, within the algebraic approach to quantum field theory, by applying the, so-called, GNS construction (by which one recovers the standard formulation of the quantum field theory in terms of a Hilbert space) to a mixed state, one arrives at a \emph{reducible} representation of the algebra of field operators. Such a feature is usually understood as indicating that states belonging to the Hilbert space of different irreducible components cannot be connected to each other because the operators capable of doing so correspond to alterations of the state of the system that are associated to inaccessible degrees of freedom. However, that interpretation would be impossible if we hold to the view that a complete description of a non-entangled system might correspond to a mixed state. In that case, it seems it would be reasonable to expect some alternative physical explanation for the above mentioned impediment.

%On a similar light, if we adopt the position advocated by Maudlin in which the state of a system on the ``accessible'' region (i.e., the connected part of the Cauchy hypersurface) might be mixed, simply because the system under description is in reality entangled with a set of degrees of freedom that lie in an inaccessible region, one would have to wonder when, if at all, a state of a system we characterize in terms of what seems accessible at the time might be regarded as pure. We could always wonder if a microscopic black holes could be generating entanglement of any system under consideration with degrees of freedom that become inaccessible. 

Given all these problems with the non-standard idea of assigning mixed states to non-entangled, isolated systems, we adhere to the standard practice in which mixed states can only occur in those situations described at the beginning of the section: proper mixtures, where the non-purity is understood in purely epistemic grounds and improper mixtures, where the lack of purity is taken to reflect the entanglement of the system with degrees of freedom external to it.

As for the usual presentation of the black hole evaporation problem, one often considers a system initially prepared in a well-known pure quantum state, which evolves, according to the unitary dynamics provided by quantum theory, into an equally pure state that, however, now contains a black hole. It is only when we decide to limit our description to those degrees of freedom lying in the black hole \emph{exterior} that we end up with a characterization of the subsystem in terms of a mixed state, which happens to have thermal characteristics. That state is, at this point, clearly an improper mixture. Can one continue to regard this mixture as an improper mixture at very late times, corresponding to the situation where the black hole has completely evaporated? Of course, one might be able to do so by taking a position where the ``interior of the black hole continues to exist'' in the sense of the picture presented by Maudlin. However, as we explained above, by doing so one has to accept a notion of ``now'' not associated, in general, with a ``where''. Therefore, one would have to say that the quantum state of the complete system is pure, but that is has parts that ``exist'' only in the enlarged notion of ``being present'' associated with a non-connected Cauchy hypersurface. It does seem one might be able to do this in self-consistent way. However, the price to pay is a radical departure from the standard usage of words and the notions that are implicitly taken as attached to those. Namely, if something exist ``now,'' we should be able to at least point towards its location.

More specifically, one would have to accept that, in any given situation, what is regard as the system of interest--- say a quantum field in what we usually consider to be a Cauchy hypersurface---might in fact have components associated with regions towards which it is not possible to point, and that, absent their inclusion and the corresponding characterization of the whole, our system must be seen as just a subsystem that might have to be described by a mixed state. In light of what we discussed above regarding the non-fundamental nature of black holes, this would be rather problematic because then, in general, we would not have a guarantee that any closed and isolated quantum state can ever be described by a pure state. It seems to us, then, that this would be an inconvenient way to talk about the situation and that, although is not internally inconsistent, it only makes it harder to discuss and focus on the questions of interest. It seems more natural to take the bull by the horns and face the fact that we are encountering a rather unprecedented situation; and to try to deal with it by maintain our ordinary way of referring to things to the maximal extent possible and then use appropriately selected wordings for those things that are extraordinary.
 
At any rate, most discussions of the black hole evaporation process simply do not dwell on the nature of the mixed states involved, so the question remains: are we supposed to be talking about proper or improper mixtures? Well, unless one has in mind the kind of picture advocated by Maudlin, one cannot take the mixed sates to refer to improper mixtures, simply because that would require identifying the remaining degrees of freedom that are presumably entangled with those of our system of interest, and such candidates are nowhere to be found (at least when sticking to the natural notion of the word ``now''). Thus, we are forced to conclude that the mixed states in question represent proper mixtures. That is, we are driven to see the mixed state as inescapably containing an epistemic component, related to the fact that the exact final state of the system is pure indeed, but unknown.\footnote{To avoid confusion, it is worth pointing out that the proper way to represent ignorance mathematically is a probability measure over pure states and not a mixed state.} A theory involving departures from unitary evolution is then necessary in order to complete the picture by (probabilistically) providing a concrete evolution for the system in question. That is, by supplying a specific unraveling of the associated Lindblad equation. It seems, then, that even though works such as \cite{Peskin} do not explicitly refer to any such theory as a natural candidate to incorporate the effective description of what must go on in the Hawking evaporation of a black hole, that is very close to what must be inferred from their position. Below we will have much more to say regarding this possible breakdown of unitarity.
%%%%%%%%%%%%%%%%%%%%%%%%%%%%%%%%%%%%%%%%%%%%%%%%%%%%%%%%%%%%%% 
\section{Black hole evaporation and the AdS/CFT conjecture}
\label{ADS-CFT}
%%%%%%%%%%%%%%%%%%%%%%%%%%%%%%%%%%%%%%%%%%%%%%%%%%%%%%%%%%%%%%
One further aspect we would like to comment on is the AdS/CFT correspondence (see for instance \cite{AdS/CFT} and references therein), which has been the primary motivation for recent increased interest in the black hole information puzzle. According to this conjecture, a theory of quantum gravity on asymptotically Anti-de Sitter spacetime is equivalent, through certain dualities, to an ordinary conformal field theory (CFT), without gravity, on a the boundary of such a spacetime. The implication of such a duality for the issue of black hole information is often presented as follows: given that the evolution in the CFT theory is always unitary, the same should hold for the corresponding dual process in the theory of quantum gravity, even if it involves the formation and evaporation of a black hole.

Note, however, that the above argument can be easily reversed to show precisely the opposite. If, as we will argue for below, one concludes that the evaporation of black holes leads to an effective breakdown of unitarity, then, via duality, one must conclude that unitarity must also break on the CFT side of the correspondence. As pointed out in \cite{UandW}, we should also keep in mind that, even if the AdS/CFT correspondence turns out to be exact, the argument in the last paragraph would only follow if, at the same time, one could establish that the correspondence is local in time, mapping early/late observables in the CFT side to early/late observables in the quantum gravity side (and assuming that, in the CFT theory, all observables associated to a Cauchy hypersurface form a complete set). Without such time order preserving feature of the correspondence, one would not be able to say that the connection between states on hypersurface 1 and hypersurface 2 on the gravity side are of the same nature (i.e., unitarily related) as that between states on corresponding hypersurfaces 1’ and 2’ on the CFT side. The fact that extreme care is required in this regard is illustrated by the fact that the boundary of the AdS spacetime where the CFT theory lives is globally hyperbolic, while the asymptoticaly AdS bulk spacetime is not. Thus, it is clear that the conjecture can only be correct if taken together with supplementary instructions regarding boundary conditions (further similar concerns have been raised in \cite{UandW})). And even if the AdS/CFT duality had such a desirable feature, it would seem feasible to accommodate the non-unitary relation between early and late states in the quantum gravity side if the process dual to the black hole formation and evaporation in the CFT side of the equivalence is complex enough and involves a sufficiently large number of degrees of freedom so as to generate a correspondingly large departure from unitary evolution.
%%%%%%%%%%%%%%%%%%%%%%%%%%%%%%%%%%%%%%%%%%%%%%%%%%%%%%%%%%%%%%
\section{Lessons for quantum gravity?}
\label{qg}
%%%%%%%%%%%%%%%%%%%%%%%%%%%%%%%%%%%%%%%%%%%%%%%%%%%%%%%%%%%%%%
It is often said that the information loss issue may hold the key for the construction of a quantum theory of gravity. However, in \cite{Mau:17}, Maudlin strongly disagrees with this expectation. In particular, he claims that, in order for an actual information loss problem to appear, one needs to introduce assumptions about the very theory one is trying to construct, so, at the end all one learns is what was put in by hand in the first place. We certainly agree with Maudlin regarding the fact that, in order to derive a true paradox, additional assumptions are required. However, we are not as pessimistic as he is regarding the prospect of utilizing the evaporating black hole scenario in order to explore paths towards quantum gravity---particularly if the extra assumptions required are reasonable and one is fully explicit about them. In fact, we find the exercise comparable with famous Gedanken experiments in physics, such as Newton's bucket or Einstein's elevator, or with the heuristic model construction of the old quantum theory.

The information loss paradox is often presented as an unavoidable consequence of well-established physics. However, as Maudlin clearly showed in \cite{Mau:17}, in the context of Figure 1 containing a singularity, there is no information loss, so there is nothing paradoxical or problematic going on. On the other hand, a theory of quantum gravity is widely expected to cure all singularities, drastically changing the panorama regarding the information loss issue (see Figure 3).\footnote {Explicit proposal of this kind, including figures characterizing the expected modifications of the classical Penrose diagrams with similar characteristics as Fig 3, have been advocated in works such as \cite{Viq1,Viq2,Ash,Hayward}. We think, however, that such figures are to be considered as heuristic at best. That is because quantum gravity is expected to heavily modify the nature of the fundamental degrees of freedom in the regions where a full quantum gravity description is required and thus spacetime concepts, such as the metric, would simply cease to make sense (in the same way that hydrodynamic notions, such as fluid velocity, cease to make sense when the level of description is that of a quantum theory of atoms and molecules or, at an even deeper level, quantum field theory).} 
\begin{figure}[h]
\centering
\includegraphics[height=7cm]{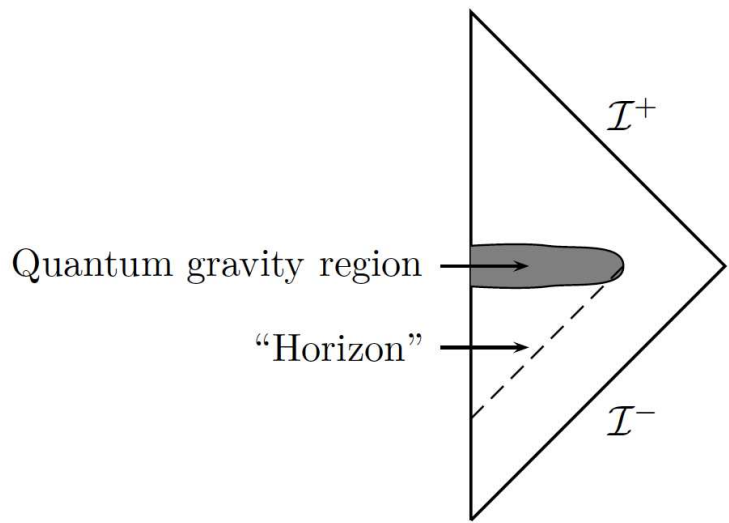} 
 \caption{``Quantum spacetime diagram'' for a black hole. The shaded area represents the region in which the relativistic description is no longer valid and the quantum gravitational one takes over.}
\end{figure}
Of course, we are not sure quantum gravity is going to actually cure the singularities, but assuming so seems to be a reasonable expectation. Therefore, exploring the consequences of such an assumption seems like a fruitful enterprise. However, just assuming that quantum gravity cures the singularity is not enough for an information loss paradox to arise. As we have argued in \cite{P13}, in order for a genuine paradox to arise, a series of non-trivial assumptions are required. In particular, we have shown that these additional assumptions involve issues regarding the nature of Hawking’s radiation, hypothesis with respect to quantum aspects of spacetime and even considerations in the foundations of quantum theory.

In more detail, the full list of assumptions we believe are necessary for an actual paradox to arise is as follows (see \cite{P13} for details):
\begin{enumerate}
\item As a result of Hawking’s radiation carrying energy away from the black hole, the mass of the black hole decreases and it either evaporates completely or leaves a small remnant.
\item In the case where the black hole leaves a small remnant, the number of its internal degrees of freedom is bounded by its mass in such a way that these cannot possibly encode the information contained in an initial state with an arbitrarily large mass.
\item Information is not transferred to a parallel universe.
\item As a result of quantum gravity effects, the internal singularities within black holes are cured and replaced by something that eliminates the need to consider internal boundaries of spacetime.
\item Information is not encoded in low-energy modes that go through the quantum gravity region.
\item The outgoing radiation associated with the region exterior to the horizon does not encode the initial information.
\item Quantum evolution is always unitary.
\end{enumerate}
In \cite{P13} we review the arguments in support of assumptions 1, 2, 3, 4 and 5 and find them reasonable. Therefore, what is left, in order to avoid a paradox, it to negate either 6 or 7. Regarding the option of negating of 6, as shown in \cite{firewalls}, it leads to a breakdown of the equivalence principle on the horizon, even if the curvature there is small and there is no local reason to expect dramatic new physics. The other option, of course, it to give up 7. In the next section we will explain why we believe this is a much more conservative and superior alternative.

%%%%%%%%%%%%%%%%%%%%%%%%%%%%%%%%%%%%%%%%%%%%%%%%%%%%%%%%%%%%%% 
\section{Breakdown of unitarity}
\label{BU}
%%%%%%%%%%%%%%%%%%%%%%%%%%%%%%%%%%%%%%%%%%%%%%%%%%%%%%%%%%%%%%
In the previous section we were explicit regarding the assumptions required in order for a true information loss paradox to arise; we saw that, in that context, a possible way out of the problem would be to give up the assumption that quantum evolution is always unitary. At the same time, it has been extensively argued that breakdown of unitarity leads to a promising solution to the measurement problem (see \cite{Bassi} for a review). Moreover, as we have shown in the last few years, such a possibility not only offers a simple path for fully diffusing the black hole information puzzle \cite{P9,P10,P11,P12,P13}, it also provides tools to solve a number of open problems in theoretical physics, such as (see \cite{weight} for review):
\begin{itemize}
\item Solving a critical conceptual problem with the inflationary account of the emergence of seeds of cosmic structure \cite{P1,P3,P6,P7}.
\item Revising the expectation that inflation will give rise to, by now detectable, B-modes in the CMB \cite{P16}.
\item Explaining dark energy in terms of energy non-conservation \cite{Dark-Energy-collapse}.
\item Dealing with the problem of time in quantum gravity \cite{P9}.
\item Shedding light on the origin of the second law of thermodynamics and the very special initial state of the universe \cite{P14}.
\end{itemize}

At any rate, what we would like to do in this section is to explain why a clue of information loss in the context of an evaporating black hole, together with a quantum gravity outlook regarding the non-fundamentality of classical spacetime, leads to the expectation of unitarity being broken, and information being lost, much more generally.

To begin with, let us think of fluid dynamics as a useful analogy to the emergence of spacetime that many researchers in quantum gravity have in mind. The description of a fluid, in terms of the Navier-Stokes equations, emerges from an underlying theory of atoms and molecules interacting via electromagnetic forces; in some regime, such forces might be described in terms of Van der Waals potentials but, in general, full-fledged quantum electrodynamics will be required. When looking at fluid dynamics in that way, it is clear that concepts such as fluid density or pressure cannot be taken to have any \emph{fundamental} character, and that phenomena such as vorticity or turbulence, which at the level of fluid mechanics are essential, cannot be expected to have any special significance or play any crucial role at the fundamental level. As a matter of fact, when looking at the issue from the standpoint of the fundamental theory, a careful analysis will reveal that, in a sense, what the fluid dynamics theoretician calls vorticity and turbulence are present, to some degree, in any real process. And, of course, the fact that, under certain conditions, they can be safely ignored, is just a matter of the degree of approximation we are interested in, and the time scale over which we will study the system. For example, one can solve the equations for a viscosity-free liquid under laminar flow conditions, say, in a circular tube, and conclude that the motion will continue forever. However, everyone acknowledges that such a result is in fact not true in nature as the tiny but ever-present viscosity effects lead to some microscopic vorticity-like behavior that ends up producing energy dissipation into heat, eventually stopping the circulatory motion.\footnote{Of course, unless one is dealing with a superfluid.} 
 
Now, when adopting the analogous view about the nature of spacetime, classical spacetime notions, such as the metric, the casual structure or Cauchy hypersurfaces are expected to become secondary or emergent. If so, it is clear that a black hole should not be considered as fundamental, any more than a hurricane is to a fluid dynamics theoretician.\footnote{The analogy between black holes and hurricanes could seem a bit strained. Unlike for a hurricane, the criteria for being a black hole are sharply defined, as is the process of black hole evaporation. In it important to remember, though, that such criteria are sharply defined only if general relativity is taken as a non-emergent theory, which is exactly the assumption we are questioning.}  To push the analogy further, suppose someone is concerned with the issue of energy conservation in fluid dynamics and considers a proposal in which energy is generally conserved, but not so when hurricanes are involved: under such conditions, the proposal holds, part of the energy becomes inaccessible. Such a suggestion will probably not be deemed satisfactory because people would be likely to suspect that, if during a hurricane some part of the energy went into a strange, inaccessible form, then similar effective losses of energy are likely to occur in other circumstances as well. People might even argue that, to some extent, small hurricanes can be though of as taking place in almost any situation involving nontrivial fluid motion. Others might argue that, in fact, at a certain scale, the concept of a hurricane becomes simply inappropriate and a description of things in another language is necessary. Regardless, one should expect an explanation in terms of the fundamental level of description of how exactly what we call energy, at the level of the fluid dynamics language, effectively disappears when a hurricane is involved. Moreover, one will also tend to expect similar processes \emph{not} involving hurricanes, also to lead to energy non-conservation, although perhaps in a less significant fashion.
 
What about black holes? If the unitary evolution is, at least in practice (as reflected, for instance, in what is available to observers that only have access to $\Sigma_{2out}$), fundamentally lost when a black hole is involved, then, to researchers that do not see spacetime as fundamental, this seems unavoidably tied with the conclusion that unitarity evolution cannot be exactly valid in any situation. That is because the very notion of a black hole is a one that is only well-defined within a setting where the emergent notion of classical spacetime becomes relevant, and thus the presence or absence of a black hole can have no implications regarding the behavior of the fundamental theory in generic situations. 

%ACA NO SE LO QUE SE QUIERE DECIR : If unitary evolution at the practical fundamental level is not the universal norm in the theory, then it cannot occur when a black hole is involved. And if it fails to hold when a black hole is involved, then it follows that such departure from unitary evolution must be generic (although perhaps in a less important qualitative level than when a black hole is involved).

Regarding the possibility of black hole evaporation breaking unitarity, early evaluations of the issue in \cite{Peskin} arrived at the conclusion that any type of evolution from pure states to mixed states would give rise to unacceptable violations of either causality or energy-momentum conservation, not only in the context of black hole evaporation, but also in ordinary laboratory situations. However, by considering simple examples of unitarity breaking dynamics, \cite{Wald-Unruh} found that, for states that might be produced in a laboratory, only unobservably small effects would arise.\footnote{In a recent private conversation, R. M. Wald pointed out that the fact that W. G. Unruh and him addressed the issue raised in \cite{Peskin}, does not mean that they accept the premises underlying such a work.} What we would like to point out is that it seems that both \cite{Peskin} and \cite{Wald-Unruh}, perhaps unwittingly, adopt a position similar to what we are advocating here. Those articles seem to tacitly acknowledge that, whatever can happen when black holes are involved, must also be considered as possibly taking place in ordinary circumstances. That is, if a departure form unitary evolution is expected when black holes evaporate, similar departures from unitary evolution can be expected in ordinary situations. Otherwise, it is hard to understand why would these works center the discussion around a possible departure from unitary evolution in everyday situations.\footnote{Let us clarify further our point of view in this regard. Imagine for a moment that that, at the fundamental level, the spacetime description is something like what is provided by loop quantum gravity. If so, at the fundamental level there is no metric and thus no notion of asymptotic flatness nor of black holes. It is clear, however, that if the theory is viable, it would have to be able to account for the emergence of a spacetime metric, and thus of black holes (i.e., that under certain circumstances, some features of the state of holonomies and fluxes would have to be identified as describing a black hole). Now, imagine that, in order to describe a physical process, one computes the amplitude using some kind of path integration (which in the case of loop quantum gravity would involve spin foams). The standard usage of path integrals involves summing over all possible intermediate paths connecting the initial and final conditions. Thus, any such complete sum over paths can be expected to contain some of the holonomies and fluxes that have been identified as corresponding to a black hole. In other words, we expect virtual black holes as part of any physical process simply because at the basic level there is no distinction between virtual and real processes.} 

%We see this concern as motivated by a view whereby excitations of quantum geometrical nature, that might be characterized as virtual black holes, can be expected to occur in all ordinary situations, and that if macroscopic black holes are taken to lead to departures from unitarity, so must these ever-present microscopic virtual black holes, and thus that departures from unitary evolution must be a common occurrence. In fact these excitations might have to be regarded as more generic than on shell black holes as they need not be either stationary, or even satisfy the laws of classical general relativity (in usual parlance these excitations need not be on shell''). 
 
At any rate, as we saw above, now we know that unitarity breaking dynamics are perfectly viable. In fact, as we mentioned above, so-called, \emph{objective collapse models}, which have been developed with foundational issues in quantum theory in mind, and which generically break unitarity, not only do not lead to predictions of problematic violations of either causality or energy-momentum conservation, but seem go be empirically adequate in all respects. Moreover, as we also mentioned, in \cite{P9,P10,P11,P12,P13} we have successfully adapted such models in order to explicitly describe the transition from a pure into a mixed state during the evaporation of a black hole. 

%As we said, the former argues that if unitarity is broken during the evaporation of a black hole, then unitarity would also be lost in general and would, according to their analysis, lead to important violations of energy conservation and causality. Similarly, the refutation in \cite{Wald-Unruh} does not worry about issues such as our distance to astrophysical black holes or the impracticality of conducing an experiment involving an evaporating black hole. Instead, the argument focuses on the departure from unitary evolution in everyday situations, in line with our position. 
 
%Some colleagues like to emphasize this point using the case of particle interactions at CERN and talk about black holes that might be produced during such events. In some quarters there is even talk of real black hole production (such as would arise for instance in the strong gravity scenario considered by Randal-Sundrum \cite{RS}) but others prefer to talk about the effect of virtual black holes. The point however is that just as with the hurricanes in our previous discussions, at the fundamental level of the theory from where the classical notions of spacetime ought to emerge, a black hole is not a fundamental object, just a coarse grained characterization of a type of object or process that is common in the underlying theory. Such objects/process would be present to a certain degree in any and all situations, and as such should be tied to what at the macro level of discussion would seem as a departure from unitary evolution.
%%%%%%%%%%%%%%%%%%%%%%%%%%%%%%%%%%%%%%%%%%%%%%%%%%%%%%%%%%%%%% 
\section{Black holes and conservation laws}
\label{con}
%%%%%%%%%%%%%%%%%%%%%%%%%%%%%%%%%%%%%%%%%%%%%%%%%%%%%%%%%%%%%%
In section \ref{now} we pointed out the fact that, during the evaporation of a black hole, violations of conservation laws for quantities such as $B$ are to be expected. In this section we would like to make a few comments on the subject. First we should note that no violation is expected to arise in connection with conservation of electric charge. This is because the electromagnetic fields in the exterior of the black hole``remember'' how much charge went into the black hole. More precisely, such fields affect the Hawking radiation of charged particles, leading to an enhanced emission of particles with the same charge as the black hole. However, regarding quantities such as $B$ and $L$, there are no similar mechanisms that could enhance the emission of, say, neutrons vs. antineutrons (due to the CPT theorem, such particles have the same mass so, in accordance with the thermal character of the Hawking radiation, they are emitted with equal rates). Thus, as noted in section \ref{now}, a black hole formed by, say, $10^{50}$ baryons is expected to lead to a final, post-evaporation state containing no net $B$.

What can be said about such a non-conservation, particularly in a context in which quantum gravity is supposed to have cured the singularity? Of course, one could ascribe it to a purely quantum gravity feature, taking place in the region that replaces the singularity (see Figure 3). However, as suggested by the discussion of section \ref{BU}, it is much more natural to assume that the non-conservation process is not restricted to the purely quantum gravity region. In fact, as mentioned in footnote \ref{spha}, both $B$ and $L$ are violated by the sphaleron mediated processes that involves the generation of a non-perturbative configuration of the boson fields (gauge bosons and Higgs) of the electroweak sector. Such a process is normally suppressed by a factor of $e^{(-E/M_{sp})}$, where $E$ is the energy of the system and $M_{sp}$ the mass of the sphaleron (which is of the order of 10 TeV). Therefore, under ordinary conditions, such violations are extremely rare, but can in fact occur without any connection with black hole physics. Now, if the non-conservation of both $B$ and $L$ is to be attributed to processes taking place throughout the interior of the black hole, and not only in the quantum gravity region, then the non-conservation process has to be substantially enhanced within the black hole. Interestingly, though, the notion of energy within the black hole is highly distorted,\footnote{That is, in the absence of gravitation, the configuration of gauge and scalar fields known as the sphaleron is characterized by a certain value of energy. However, such a value of energy is intimately tied with the symmetries of Minkowski spacetime. Within a black hole, and particularly an evaporating one, the notion of energy is not even well-defied. Therefore, the configuration of the filed corresponding to the sphaleron would, quite likely, be rather different from its flat spacetime counterpart. As a result, under such conditions it is unclear exactly how the energetic suppression of the sphaleron mediated processes would manifest itself.}  in which case it is natural to expect otherwise highly suppressed processes to occur much more readily within the black hole. In particular, one could expect an enhanced the rate of the $B$ and $L$ violation process, perhaps even allowing it to account for the complete elimination of any $B$ or $L$ net number in the black hole interior before the quantum gravity region is reached.

On the other hand, given the fact that, within the standard model of particle physics, there is no mechanism which could violate the conservation of $B-L$, it is clear that the picture we have presented above cannot be used to address the non-conservation of $B-L$ through the evaporation of a black hole. Therefore, if we want to develop a detailed proposal in line with what we proposed in section \ref{BU}, namely, in which processes within black holes are to occur as well, albeit in a highly suppressed way, in everyday situations, then, in order to fully account for the violation of laws such as the conservation of $B-L$, we have to assume that such violations take place in ordinary processes. In particular, if, as in \cite{P9,P10,P11,P12,P13}, one is advocating an approach to these questions based on an objective collapse model, then the collapse mechanism should also incorporate a direct violation of $B-L$, which must take place, perhaps at highly reduced rates, in ordinary situations. In other words, if objective collapse models are to successfully explain the evaporation process, then they should violate $B-L$.\footnote{For quantitative reasons, the theory should also lead to enhanced violation of T and CP symmetries, known occur in the standard model of particle physics only in a highly suppressed manner.} We believe this conclusion clearly illustrates the potential for insights about the nature of novel aspects of physics that can be extracted from considerations regarding the formation and evaporation of black holes.

%%%%%%%%%%%%%%%%%%%%%%%%%%%%%%%%%%%%%%%%%%%%%%%%%%%%%%%%%%%%%%
\section{Conclusions}
\label{C}
%%%%%%%%%%%%%%%%%%%%%%%%%%%%%%%%%%%%%%%%%%%%%%%%%%%%%%%%%%%%%%
Since the publication of Hawking's analysis, more than forty years ago, the issue of black hole information loss has remained a central topic of analysis and debate in theoretical physics. The AdS/CFT correspondence, proposed almost twenty years latter, reignited the notorious debate, but unfortunately did not help in clarifying its terms. The fact is that, after all this time, the discussion continues, often hindered and exacerbated by confusion and misunderstanding among participants. We hope the present article helps clarify some aspect of the discussion and highlight some of the attractive features of our own take on the subject. 
%%%%%%%%%%%%%%%%%%%%%%%%%%%%%%%%%%%%%%%%%%%%%%%%%%%%%%%%%%%%%% 
%%%%%%%%%%%%%%%%%%%%%%%%%%%%%%%%%%%%%%%%%%%%%%%%%%%%%%%%%%%%%%
\section*{Acknowledgments}
%%%%%%%%%%%%%%%%%%%%%%%%%%%%%%%%%%%%%%%%%%%%%%%%%%%%%%%%%%%%%%
We thank T. Maudlin and R. M. Wald for very useful discussions. We acknowledge partial financial support from DGAPA-UNAM project IG100316. DS was further supported by CONACyT project 101712.
%%%%%%%%%%%%%%%%%%%%%%%%%%%%%%%%%%%%%%%%%%%%%%%%%%%%%%%%%%%%%%
%%%%%%%%%%%%%%%%%%%%%%%%%%%%%%%%%%%%%%%%%%%%%%%%%%%%%%%%%%%%%%
%%%%%%%%%%%%%%%%%%%%%%%%%%%%%%%%%%%%%%%%%%%%%%%%%%%%%%%%%%%%%%
\bibliographystyle{plain}
\bibliography{biblio}
%%%%%%%%%%%%%%%%%%%%%%%%%%%%%%%%%%%%%%%%%%%%%%%%%%%%%%%%%%%%%%
\end{document}